\documentclass[12pt]{iopart}   
\begin{document}

\title[Random walks in a random environment on a strip]{Random walks in a random environment on a strip: a renormalization group approach}

\author{R\'obert Juh\'asz} 
\address{Research Institute for Solid
State Physics and Optics, H-1525 Budapest, P.O.Box 49, Hungary}
\ead{juhasz@szfki.hu} 

\begin{abstract}
We present a real space renormalization group scheme 
for the problem of random walks in a random environment on a
strip, which includes the one-dimensional random walk 
in random environment with bounded non-nearest-neighbour jumps. 
We show that the model
renormalizes to an effective one-dimensional random walk with
nearest-neighbour jumps and
conclude that Sinai scaling is valid in the recurrent case, while
in the sub-linear transient phase, the displacement grows as a
power of the time.   
\end{abstract}

\submitto{J. Phys. A: Math. Theor.}
\maketitle

\newcommand{\bc}{\begin{center}}
\newcommand{\ec}{\end{center}}
\newcommand{\be}{\begin{equation}}
\newcommand{\ee}{\end{equation}}
\newcommand{\beqn}{\begin{eqnarray}}
\newcommand{\eeqn}{\end{eqnarray}}

\section{Introduction}

The problem of random walks in a random environment (RWRE) has a long
history and since the early results in the 1970s
\cite{sk}, a vast amount of informations have
accumulated; for a recent review see Ref. \cite{zeitouni}.  
The RWRE can be regarded as a toy model of disordered systems, 
for which exact results
are available and which, due to its simple formulation, became a 
fundamental model in various fields such as transport processes or
statistical mechanics of magnetic systems \cite{ab}.   
Most works concern the RWRE with nearest-neighbour jumps
on the integers, for which a more or less complete picture is at our
disposal. 
Beside rigorous results \cite{sk,sinai,golosov}, 
this model was also studied by a strong
disorder renormalization group (SDRG) method \cite{dmf} which is
closely related to that originally developed for disordered spin models \cite{mdh}. 
This method, in which the small barriers of the energy  
landscape are successively eliminated, yields exact results for the asymptotical
dynamics, among others the scaling of the typical displacement $x$ of the
walker with time $t$ in the recurrent case: $x\sim(\ln t)^2$, in 
accordance with Sinai's theorem \cite{sinai}.       

In higher dimensions, even on quasi-one-dimensional lattices or in case
of non-nearest-neighbour jumps, the understanding of RWRE is at present
far from complete. 
For the one-dimensional(1D) RWRE with bounded non-nearest-neighbour jumps, 
criteria for recurrence and
transience are known \cite{key} and for some special cases 
Sinai scaling was proven \cite{lb}. This model arises
also in the context of disordered dynamical systems \cite{radons}. 
For the RWRE on strips of finite width, which
incorporates among others the
former model and the persistent RWRE \cite{szta},  recurrence and
transience criteria were obtained in Ref. \cite{bolthausen}.  
  
The aim of this paper is to propose an exact SDRG scheme 
for the RWRE on a strip. 
A necessary condition for the analytical tractability by the SDRG method is
that the topology of the underlying lattice is invariant under the
transformation, which generally does not hold apart from 1D. 
As in our approach complete layers of lattice sites are decimated,
the topology of the network of transitions is preserved.   
Contrary to the 1D RWRE, the energy landscape does not
exist in general, therefore we keep track of the transformation of jump
rates in the same spirit as it was done for the closely related
1D zero range process \cite{jsi}.   
We shall show that in the
fixed point, the transformation of relevant variables is identical to
that of the 1D RWRE with nearest-neighbour jumps, 
implying among others that Sinai scaling 
holds for strips of finite width in the recurrent case.   

The rest of the paper is organised as follows. In Section
\ref{secform}, the problem to be studied is defined in details. 
In Section \ref{secrg}, the renormalization group (RG) transformation is
introduced and the RG equations are analysed in the recurrent case, as
well as in the zero-velocity transient phase. 
Finally, the results are discussed in Section \ref{secdisc}. 

\section{Formulation of the problem}
\label{secform}
 
We consider a finite strip $S=\{1,\dots,L\}\times\{1,\dots,m\}$ of
length $L$ and width $m$, and call the set of sites $(n,i)\in S$ with
fixed $n$ and $i=1,\dots,m$ the $n$th layer.  
We define on this lattice a continuous-time random walk
by the following (nonnegative) transition rates for 
$1\le n\le L$:
\beqn
T(z_1,z_2)=\left\{
\begin{array}{ccc}
P_n(i,j) &{\rm if}& z_1=(n,i), z_2=(n+1,j)  \nonumber \\
Q_n(i,j) &{\rm if}& z_1=(n,i), z_2=(n-1,j) \nonumber \\
R_n(i,j) &{\rm if}& z_1=(n,i), z_2=(n,j), i\neq j \nonumber \\
0        & & {\rm otherwise}. \nonumber 
\end{array}
\right. 
\label{rates}
\eeqn
Here and in the following, 
the formally appearing index $(0,j)$ [$(L+1,j)$] is meant to
refer to site $(L,j)$ [$(1,j)$],  
i.e. the strip is periodic in the first coordinate. 
The $m\times m$ matrix $P_n$($Q_n$) contains the jump rates from
the $n$th layer to the adjacent layer on the right(left), while the
matrix $R_n$ with diagonal elements 
$R_n(i,i):=-\sum_{j\neq i}R_n(i,j)$ contains the intra-layer jump rates.  
Besides, we define the $m\times m$ matrix $S_n$, which will be useful
in later calculations by 
$S_n(i,j):=-R_n(i,j)$, $i\neq j$, while  the
diagonal elements are fixed by 
\be 
(P_n+Q_n-S_n){\bf 1}={\bf 0},
\label{stoch}
\ee 
where ${\bf 1}$(${\bf 0}$) is a column vector with all components $1$($0$).   
For the sequence of triples of matrixes, $\{(P_n,Q_n,R_n)\}$, which defines
the random environment, we impose at this point the only condition that it must
be connected in the sense that every site is reachable
from every other site through sequences of consecutive transitions 
with positive rates. 
The probability that the walker resides on site
$(n,i)$ in the stationary state is denoted by $\pi_n(i)$ and these are
normalized as $\sum_{(n,i)}\pi_n(i)=1$.  
Following Ref. \cite{bolthausen}, we introduce the row vectors
$\pi_n=(\pi_n(i))_{1\le i\le m}$ and for a fixed environment, 
write the system of linear equations that the stationary
probabilities satisfy in the form:
\be
\pi_nS_n=\pi_{n-1}P_{n-1}+\pi_{n+1}Q_{n+1}, \quad 1\le n\le L.
\label{system}
\ee
Although, we started from a continuous-time random
walk, the same equations can be written for a discrete-time
jump process with transition probabilities obtained by rescaling the 
transition rates by $\max_{(n,i)}S_n(i,i)$. 

\section{Renormalization group transformation}
\label{secrg}

The elementary step of the renormalization group 
method we apply is the elimination
of the $k$th layer, such that the walker then jumps from the $k-1$st layer
directly to the $k+1$st one with transition rates $\tilde P_{k-1}(i,j)$
and from the $k+1$st layer to the $k-1$st one with rates $\tilde
Q_{k+1}(i,j)$. We choose the matrices $\tilde P_{k-1}$ and
$\tilde Q_{k+1}$ in such a way that the remaining $L-1$ equations in
(\ref{system}) are fulfilled by 
the unchanged vectors $\pi_n$, $n\neq k$. 
Eliminating $\pi_k$ in Eq. (\ref{system}), it turns out
that also the matrices $S_{k-1}$ and $S_{k+1}$ must be changed, 
and we have the following transformation rules:
\beqn
\tilde P_{k-1}&=&P_{k-1}S_k^{-1}P_k   
\label{rule1} \\
\tilde Q_{k+1}&=&Q_{k+1}S_k^{-1}Q_k    
\label{rule2}    \\
\tilde S_{k-1}&=&S_{k-1}-P_{k-1}S_k^{-1}Q_k  
\label{rule3}    \\
\tilde S_{k+1}&=&S_{k+1}-Q_{k+1}S_k^{-1}P_k.          
\label{rule4}  
\eeqn
All other matrices remain unchanged.  
The matrix $S_n$ has the following important property: 
\be
S_n^{-1} \ge 0, 
\label{invpoz} 
\ee
which is meant to hold for the matrix elements. 
This can be proven as follows. 
We introduce the notation $D_m\equiv\det S_n$ where the 
index $m$ refers to the order of the matrix. The non-diagonal
elements of $S_n$ are nonpositive, while 
$S_n(i,i)\equiv\sum_j[P_n(i,j)+Q_n(i,j)]+\sum_{j\neq i}R_n(i,j)>0$ for
$1\le i\le m$ since by assumption, the environment is connected. 
Regarding $D_m$ as a function of the variables 
$\epsilon_i:=\sum_jS_n(i,j)=\sum_j[P_n(i,j)+Q_n(i,j)]$,
i.e. $D_m=D_m(\epsilon_1,\dots,\epsilon_m)$,  
it is clear that $D_m(0,\dots,0)=0$ and  
$\frac{\partial D_m}{\partial\epsilon_i}=D_{m-1}^{(i)}$
where
$D_{m-1}^{(i)}$ is the determinant of the matrix $S_n^{(i)}$ 
obtained from $S_n$
by deleting the $i$th row and column. 
Now, the relation $D_m>0$ can be shown by
induction. Obviously, $D_1>1$. Assuming that 
$D_{m-1}^{(i)}=\frac{\partial D_m}{\partial\epsilon_i}>0$ for
$1\le i\le m$ and
taking into account that connectedness implies $\sum_i\epsilon_i>0$,
it follows that $D_m>0$.   
Thus $\det S_n$, as well as the diagonal elements of $S_n^{-1}$ are
positive. 
Using this result, the relations $S_n^{-1}(i,j) \ge 0$ for $i\neq j$  
can be shown again by induction in a straightforward way. 

Relation (\ref{invpoz}) and Eq. (\ref{rule3}) imply that 
$\Delta S_{k-1}\equiv \tilde S_{k-1}-S_{k-1}=-P_{k-1}S_k^{-1}Q_k\le
0$. In components: 
\be
\Delta R_{k-1}(i,j)\ge 0 \quad  (i\neq j), \qquad 
\Delta S_{k-1}(i,i) \le 0.
\nonumber
\ee
From these relations we obtain 
$\sum_j\Delta P_{k-1}(i,j) \le 0$, where we have used $\Delta Q_{k-1}=0$.
Similarly, we obtain: $\Delta R_{k+1}(i,j)\ge 0$, $i\neq j$ and 
$\sum_j\Delta Q_{k+1}(i,j) \le 0$. 
Thus, the intra-layer transition rates are non-decreasing, while 
the sum of rates of inter-layer jumps starting from a given site
is non-increasing under a renormalization step.

Let us introduce the quantity $\Omega_n:=1/\|S_n^{-1}\|$, where the matrix
norm $\|\cdot\|$ is defined as $\|A\|:=\max_{i}\sum_j|A(i,j)|$. 
From Eq. (\ref{rule3}), we have 
$\tilde S_{k-1}^{-1}=S_{k-1}^{-1}+
S_{k-1}^{-1}P_{k-1}S_k^{-1}Q_k\tilde S_{k-1}^{-1}$. 
As relation (\ref{invpoz}) is valid also for the renormalized
matrices, i.e. $\tilde S_{k-1}^{-1},\tilde S_{k+1}^{-1}\ge 0$, 
both terms on the right hand side are nonnegative, therefore 
$\|\tilde S_{k-1}^{-1}\|=\|S_{k-1}^{-1}+S_{k-1}^{-1}P_{k-1}S_k^{-1}Q_k\tilde S_{k-1}^{-1}\|
\ge \|S_{k-1}^{-1}\|$, or, equivalently,  
$\tilde \Omega_{k-1} \le \Omega_{k-1}$. 
By a similar calculation we
obtain that $\tilde \Omega_{k+1} \le \Omega_{k+1}$. 
The RG procedure for finite $L$ is defined as follows. 
The layer with the actually largest $\Omega_n$ is decimated, which
results in a RWRE on a one layer shorter strip with effective rates
given by Eqs. (\ref{rule1}-\ref{rule4}) and the remaining $\pi_n$
unchanged. This step is then iterated until a single layer is left. 
The variable defined by $\Omega:=\max_n\Omega_n$, where $n$ runs
through the set of indices of non-decimated (or active) layers, 
decreases monotonously in the course of the procedure.  
For the special case $m=1$ (1D), 
$\Omega_k=Q_k(1,1)+P_k(1,1)$ and the transformation rules 
reduce to 
\beqn
\tilde P_{k-1}(1,1)&=&\frac{P_{k-1}(1,1)P_k(1,1)}{Q_k(1,1)+P_k(1,1)},  
\nonumber \\
\tilde Q_{k+1}(1,1)&=&\frac{Q_{k+1}(1,1)Q_k(1,1)}{Q_k(1,1)+P_k(1,1)},
\label{1d}
\eeqn
 which have already been obtained in the context of the zero range
 process \cite{jsi}. 

The procedure described so far applies to any connected environment; 
as a trivial case even to the homogeneous environment. 
From now on we assume that the triples ($P_n,Q_n,R_n$)
are independent, identically distributed random variables. 
We consider an infinite sequence of triples
$\{ (P_n,Q_n,R_n)\}$ and, in the usual continuum
formulation \cite{fisher} of the above RG procedure, 
we are interested in the asymptotic scaling of $\Omega$ 
with the length scale $\xi_{\Omega}$ that is given by the inverse of the
number density $c_{\Omega}$ of active layers: $\xi_{\Omega}\equiv 1/c_{\Omega}$.

\subsection{Recurrent case}

First, we focus on the case of transition rate distributions 
for which the random walk is recurrent in almost every environment. 
The question of recurrence is in general non-trivial 
for $m>1$ \cite{key,bolthausen}; 
nevertheless, a sufficient condition of recurrence is that 
the distribution of jump rates is invariant under the interchange of
$P_n$ and $Q_n$ \cite{kr}. 
Furthermore, we do not deal with special environments 
which lead to normal diffusive behaviour (e.g. the case $P_n=Q_n$ for all $n$).
Instead, we consider less restricted situations: for instance,
distributions where $P_n$ and $Q_n$ are independent. In this case, the above special
environments form only a zero-measure set in the limit $L\to\infty$.    

As a first step, we investigate the limits of transition rates 
when the density of active layers $c_{\Omega}$ goes to zero. 
Consider a site $(n,i)$ in an active layer in an arbitrary stadium of
the RG procedure and assume that the initial matrix elements $S_n(i,j)$
were renormalized to some $\tilde S_n(i,j)\le S_n(i,j)$. 
Then we can write 
$\sum_{j\neq i}\tilde R_n(i,j) \le \sum_{j\neq i}\tilde R_n(i,j)   
+\sum_j[\tilde P_n(i,j)+\tilde Q_n(i,j)]\equiv 
\tilde S_n(i,i)\le S_n(i,i)$. 
Consequently, the intra-layer rates remain bounded 
throughout the RG procedure. 
Writing, e.g., eq. (\ref{rule3}) in the form 
$\Delta S_{k-1}=-P_{k-1}S_k^{-1}Q_k$, we see that at least one of
the sets of matrices $\{P_n\}$ and $\{Q_n\}$ must tend to zero as
$c_{\Omega}\to 0$, otherwise the matrices $S_n$ would not remain
bounded. Furthermore, it is clear that the assumption 
on recurrence requires that both $\{P_n\}$ and $\{Q_n\}$ must tend 
to zero if $c_{\Omega}\to 0$. 
This also implies that, in that limit, $\det S_n \to 0$ and 
$\Omega \to \Omega^*=0$.
So, as the RG transformation progresses the inter-layer rates at the
non-decimated layers are approaching zero without limits.  

For the study of various quantities close to the fixed point
$\Omega^*=0$, it is expedient to define the following relation: 
$f\simeq g$ if $\lim_{\Omega\to 0}f/g=1$.  
According to the above, we have 
$\tilde S_{k-1}\simeq S_{k-1}$ and similarly, for the matrix  
$\mathcal{S}^{-1}_n:=S_n^{-1}/\|S_n^{-1}\|$, 
$\tilde {\mathcal{S}}^{-1}_{k-1}\simeq {\mathcal{S}}^{-1}_{k-1}$ holds. 
One can easily show that
the rows of $\tilde{\mathcal{S}}^{-1}_n$ are asymptotically identical, 
i.e. $\tilde{\mathcal{S}}^{-1}_n(i,j)\simeq\tilde{\mathcal{S}}^{-1}_n(k,j)$
for $1\le i,j,k\le m$, and the vectors formed from the rows tend to
the stationary measure $\tilde \pi_n$ of the isolated $n$th
layer, i.e. 
$\tilde {\mathcal{S}}^{-1}_n(i,j)\simeq \tilde \pi_n(j)$ 
for $1\le i,j\le m$, where 
$\tilde \pi_n$ is the solution of the equation 
$\tilde\pi_n\tilde R_n=0$ which fulfils the
condition $\sum_i\tilde\pi_n(i)=1$. 
Although, the layers were not assumed to be connected within themselves
initially, after many decimations they become almost
surely connected due to the generated positive intra-layer transition
rates when eliminating adjacent layers. If it is the case, the measure 
$\tilde\pi_n$ is unique. 
Introducing the matrices $\mathcal{P}_n:={\mathcal{S}}^{-1}_nP_n$
and $\mathcal{Q}_n:={\mathcal{S}}^{-1}_nQ_n$, Eq. (\ref{rule1}) 
can be written as 
$\tilde{\mathcal{P}}_{k-1}-\tilde
P_{k-1}\Delta_{k-1}=\mathcal{P}_{k-1}\mathcal{P}_{k}/\Omega_k$ with 
$\Delta_k\equiv\tilde {\mathcal{S}}^{-1}_{k-1}-{\mathcal{S}}^{-1}_{k-1}$. 
Using Eq. (\ref{stoch}) we obtain that 
$\|S_k^{-1}(P_k+Q_k)\|=1$. 
The rows of $S_k^{-1}$ are asymptotically identical, therefore 
$\|S_k^{-1}P_k\|+\|S_k^{-1}Q_k\|\simeq \|S_k^{-1}(P_k+Q_k)\|=1$ and 
$\Omega_k\simeq \|\mathcal{P}_k\|+\|\mathcal{Q}_k\|$.  
Furthermore, $\Delta_k\to 0$ if $\Omega\to 0$, 
thus we obtain the asymptotical renormalization rule
$\tilde{\mathcal{P}}_{k-1}\simeq
{\mathcal{P}}_{k-1}{\mathcal{P}}_{k}/(\|\mathcal{P}_k\|+\|\mathcal{Q}_k\|)$,
and we have a similar equation for $\tilde{\mathcal{Q}}_{k+1}$. 
Using that the rows of $\|\mathcal{P}_k\|$ are
asymptotically identical, we have 
$\|{\mathcal{P}}_{k-1}{\mathcal{P}}_{k}\|\simeq
\|{\mathcal{P}}_{k-1}\|\cdot\|{\mathcal{P}}_{k}\|$ and obtain finally: 
\be
\|\tilde{\mathcal{P}}_{k-1}\|\simeq\frac{\|{\mathcal{P}}_{k-1}\|\cdot\|{\mathcal{P}}_{k}\|}{\|\mathcal{P}_k\|+\|\mathcal{Q}_k\|},
\quad 
\|\tilde{\mathcal{Q}}_{k+1}\|\simeq\frac{\|{\mathcal{Q}}_{k+1}\|\cdot\|{\mathcal{Q}}_{k}\|}{\|\mathcal{P}_k\|+\|\mathcal{Q}_k\|}.
\label{rg1}
\ee
We see that these equations have the same form as those of the
1D RWRE in Eq. (\ref{1d}). 
The physical interpretation of these results is clear. If $\Omega\ll
1$, the effective inter-layer rates 
are much smaller than the effective intra-layer rates, thus
the walker in the renormalized environment spends very long time 
in a layer until it jumps to another
one, so that its quasistationary distribution within the layer is given
asymptotically by $\tilde\pi_n$. When the walker leaves the layer it
does not ``remember'' at which site it entered the layer and
irrespectively of this site, the effective jump rates to the 
adjacent layer to the right and left are 
$\|\tilde{\mathcal{P}}_{n}\|$ and  $\|\tilde{\mathcal{Q}}_{n}\|$,
respectively.   
Thus we may say that the model under study renormalizes 
asymptotically to a 1D RWRE. 
In the course of the RG transformation, 
the normalization of the measure is obviously not conserved, i.e. 
$\sum'_{(n,i)}\pi_n(i)<1$, where the prime denotes that 
the summation goes over the active sites.
Nevertheless, on a finite strip, the walker spends most of the time
in a small number of layers and the sum of $\pi_n(i)$
over almost all sites goes to zero in the limit $L\to\infty$, 
which is closely related to the Golosov localization \cite{golosov}.
At any stage of the RG transformation, the layer with the maximal 
$\Omega_n$ is decimated and $\Omega_n\sum_i\pi_n(i)$  
can be interpreted, at least close to the fixed point, as the probability
current from the $n$th layer to the neighbouring ones. 
This ensures that layers with smaller
$\sum_i\pi_n(i)$, i.e. where the walker can be found with a smaller
probability, are decimated typically earlier in the course of 
the SDRG procedure. 
Thus, fixing the length scale $\xi>1$ and renormalizing a finite 
strip of length $L>\xi$ to a strip of length $L'= L/\xi$, 
we expect that $\sum'_{(n,i)}\pi_n(i)\to \mathcal{O}(1)$ 
almost always if $L\to\infty$. Now, if the correct normalization of
$\pi_n(i)$ in the renormalized strip is restored by 
dividing by $\sum'_{(n,i)}\pi_n(i)$,
the probability current along the strip is modified only by 
an $\mathcal{O}(1)$ factor.  
On the other hand, the current is invariant under the RG transformation, thus 
assuming that $\xi\gg 1$, the RWRE on a strip
of length $L$ has the same current up to an $\mathcal{O}(1)$ factor
as an effective 1D RWRE of length $L'\sim L$. 
This implies that the current of the RWRE on a strip must
asymptotically scale with the size as that of the 1D RWRE. 
Consequently, the inverse of the current, which gives 
the mean time $\tau$ that the walker needs to make a complete
tour on the strip, must scale with $L$ asymptotically 
just as in one dimension:
\be 
(\ln \tau)^{2}\sim L.
\label{Lscaling}
\ee

Next, we have a closer look on the RG equations (\ref{rg1}) and
determine the scaling relation between $\Omega$ and 
$\xi_{\Omega}$  by pointing out the asymptotic equivalence to an 
already solved problem.  
In order to do this, we assume that the distributions of effective
rates $\|{\mathcal{P}}\|$ and  $\|{\mathcal{Q}}\|$ 
broaden on logarithmic scale without limits as $\Omega\to 0$.
This property, which can be justified {\it a posteriori}, 
is characteristic of the so-called infinite randomness
fixed points and ensures the asymptotical exactness of the
procedure \cite{fisher}. 
As a consequence, at the layer to be decimated, almost surely 
either $\|{\mathcal{P}}_{k}\|/\|{\mathcal{Q}}_{k}\|$ or
$\|{\mathcal{Q}}_{k}\|/\|{\mathcal{P}}_{k}\|$ tends to zero if $\Omega\to 0$. 
In the first case, 
$\Omega\simeq\|{\mathcal{P}}_{k}\|+\|{\mathcal{Q}}_{k}\|\simeq
\|{\mathcal{Q}}_{k}\|$ and 
the decimation rules read 
\be 
\|\tilde{\mathcal{P}}_{k-1}\|\simeq\frac{\|{\mathcal{P}}_{k-1}\|\cdot\|{\mathcal{P}}_{k}\|}{\|\mathcal{Q}_k\|}, 
\qquad \|\tilde{\mathcal{Q}}_{k+1}\|\simeq \|{\mathcal{Q}}_{k+1}\|,
\label{rg2a}
\ee
while in the second case $\Omega\simeq\|{\mathcal{P}}_{k}\|$ and 
\be 
\|\tilde{\mathcal{P}}_{k-1}\|\simeq \|{\mathcal{P}}_{k-1}\|,
\qquad
\|\tilde{\mathcal{Q}}_{k+1}\|\simeq\frac{\|{\mathcal{Q}}_{k+1}\|\cdot\|{\mathcal{P}}_{k}\|}{\|\mathcal{P}_k\|}.
\label{rg2b}
\ee 
For the above transformation rules, 
it has been shown in Ref. \cite{fisher} in the continuum limit 
that the distributions of  
$\|{\mathcal{P}}\|$ and $\|{\mathcal{Q}}\|$ flow in the recurrent case
(apart from some singular initial distributions) to the
strongly attractive self-dual fixed point with identical distribution
of $\|{\mathcal{P}}\|$ and $\|{\mathcal{Q}}\|$:
$\rho^*(\eta)=e^{-\eta}\Theta(\eta)$, where 
$\eta\equiv \ln(\Omega/\|{\mathcal{P}}\|)/\ln(\Omega_0/\Omega)$,
$\Omega_0$ is the initial value of $\Omega$ and $\Theta(x)$ is the
Heaviside step function. Furthermore, the asymptotic scaling relation
between $\xi_{\Omega}$ and $\Omega$ reads:
\be
\xi_{\Omega}\sim \ln^{2}\left(\Omega_0/\Omega\right).
\label{rgscaling}
\ee
Carrying out the RG transformation in a finite but long strip up to
the last layer which is indexed by $l$, 
the magnitude of the current can be written as
$|J|=|\pi_l(\tilde P_l-\tilde Q_l)|\approx
\sum_i\pi_l(i)|(\|\mathcal{P}_l\|-\|\mathcal{Q}_l\|)|\sim 
\sum_i\pi_l(i)\Omega_l$, where we used in the last step that for large
$L$, $\|\mathcal{P}_l\|$ and $\|\mathcal{Q}_l\|$ differ typically by
many orders of magnitude. Taking into account that $\sum_i\pi_l(i)$ is
expected to remain finite for almost all environments in the limit
$L\to\infty$ and substituting $L$ for the length scale in
Eq. (\ref{rgscaling}) we arrive again at Eq. (\ref{Lscaling}). 
From this scaling relation we conclude that the 
typical displacement of the first coordinate $x$ of the walker on 
an infinite strip scales with the time in the recurrent case as 
$x\sim (\ln t)^{2}$ for almost all environments.

\subsection{Sub-linear transient phase}

Now, we consider the case when the environment is still an
independent, identically distributed sequence but the
random walk is transient. It is known for the
1D RWRE that if $0<\mu_1 <1$, where $\mu_1$ is the unique
positive root of the equation 
$\overline{\left[Q(1,1)/P(1,1)\right]^{\mu_1}}=1$ and
the over-bar denotes averaging over the distributions 
of $Q(1,1)$ and $P(1,1)$, 
the displacement grows sub-linearly as $x\sim t^{\mu_1}$ \cite{sk,derrida}. 
In the analogous zero-velocity transient phase of the RWRE on a strip,
the matrices $P_n$ and $Q_n$ must still renormalize to zero, and the
asymptotical transformation rules are given by
Eqs. (\ref{rg2a}-\ref{rg2b}). 
The analysis of these RG equations in the continuum
limit has been carried out in Ref. \cite{ijl} and 
has yielded the asymptotical result:
$\xi_{\Omega}\sim (\Omega/\Omega_0)^{-\mu}$.
We thus conclude that the displacement grows as $x\sim t^{\mu}$ 
also for the RWRE on a strip in this phase.  
For the 1D RWRE, $\mu=\mu_1$, which is due to the fact
that the energy landscape defined by
$U_{n+1}-U_{n}=\ln[Q_{n+1}(1,1)/P_n(1,1)]$ carries the
full information on $\mu_1$ and even the approximative rules in 
Eqs. (\ref{rg2a}-\ref{rg2b}) leave the energy difference between
active sites invariant (cf. the method in Ref. \cite{dmf}). 
For $m>1$, Eqs. (\ref{rg2a}-\ref{rg2b}) are valid only asymptotically 
and the problem how the exponent $\mu$ is related to the 
initial distribution of jump rates is out of the scope of this approach.

\section{Discussion}
\label{secdisc}

We have presented in this work an SDRG scheme 
for the RWRE on quasi-one-dimensional lattices, 
which incorporates also the RWRE with bounded non-nearest neighbour jumps. 
We have made use of that by eliminating
appropriately chosen groups of lattice sites, the topology of the
network of transitions remains invariant. 
We mention that there are special sub-networks of transitions 
with positive rates which are invariant under the transformation: 
As can be seen from Eqs. (\ref{rule1}-\ref{rule2}), 
if the $i$th row or column
of $P_n$ or $Q_n$ is zero for all $n$, then this remains valid also
after an RG step. 
An example for $m=2$ is the process with the only positive inter-layer rates 
$P_n(1,1)$ and $Q_n(2,2)$, which can be interpreted as a 1D persistent
RWRE. 
We have shown that the model renormalizes to an effective
1D RWRE and concluded that, although, the finite-size corrections are
strong (see Ref. \cite{radons}), Sinai scaling is valid asymptotically in the 
recurrent case, while in the sub-linear transient
regime the displacement grows as $x\sim t^{\mu}$. 
Although, the method is not appropriate for establishing an analytical
relation between the non-universal exponent $\mu$ and the
initial distribution of jump rates,
the numerical implementation of the exact RG
scheme provides a more efficient tool for the estimation of
$\mu$ than the direct solution of Eqs. (\ref{system}).

When this work was finalized, a preprint by Bolthausen and Goldsheid
appeared, in which similar results are obtained in the recurrent
case in a different way \cite{bg}.


\ack
This work has been supported by the National Office of Research and 
Technology under grant no ASEP1111.


\end{document}